\documentstyle[prl,aps,epsf,times,float,graphicx]{revtex}
\graphicspath{{/home/janko/AAA/}}
\DeclareGraphicsExtensions{.eps,.ps}
\begin{document}
\draft
\twocolumn[\hsize\textwidth\columnwidth\hsize\csname@twocolumnfalse\endcsname
\author{A.A. Abrikosov}
\title{Theory of High-Tc Superconducting Cuprates Based on
Experimental Evidence}
\address{Materials Science Division, Argonne National Laboratory \\
9700 South Cass Avenue, Argonne, Illinois 60439, USA} 
\date{\today} \maketitle
\begin{abstract}
A model of superconductivity in layered high-temperature
superconducting cuprates is proposed, based on the extended saddle
point singularities in the electron spectrum, weak screening of the
Coulomb interaction and phonon-mediated interaction between electrons
plus a small short -range repulsion of Hund's, or spin-fluctuation,
origin. This permits to explain the large values of $Tc$, features of
the isotope effect on oxygen and copper, the existence of two types of
the order parameter, the peak in the inelastic neutron scattering, the
positive curvature of the upper critical field, as function of
temperature etc.
\end{abstract}
\pacs{PACS numbers:74.20.Fg,74.25.Kc,74.62.Dh,74.72.Bk}
]
\makeatletter
\global\@specialpagefalse
\def\@oddhead{REV\TeX{} 3.0\hfill MSD-ANL Theory Group Preprint}
\let\@evenhead\@oddhead
\makeatother

\section{Phonon-mediated interaction}

The origin and the symmetry of the order parameter in the HTSC is the
primary topic in the physics of high-temperature layered cuprates. The
crystalline symmetry, even if it is assumed tetragonal, and the order
parameter to be real (this follows from time-reversal symmetry),
requires only that at a $\pi/2$ rotation in the ab-plane it either
does not change, or changes its sign. Symmetry by itself does not tell
us how many nodes must exist in either case, or whether the formula
$\cos k_x - \cos k_y$ is correct. Therefore the only way to find the
order parameter is to study the interaction between electrons, which
leads to pairing. The high values of Tc and absence of the isotope
effect in optimally doped $YBa_2Cu_3O_{7-\delta}$ (YBCO) led to the
conclusion that this interaction cannot be mediated by phonons.
However the experiments by the groups of J. Franck and D. Morris
\cite{[1],[2]} showed that a partial substitution of $Y$ by $Pr$, or
of $Ba$ by $La$ leads to the increase of the isotope effect
simultaneously with the decrease of the critical temperature, so that
at $Tc \approx 40 K \, \alpha \approx 0.4 \, (T_c \propto M^{-\alpha})$
(Fig.1). An alternative appears: either these substituted compounds
have a different mechanism of superconductivity compared to YBCO with
$Tc \approx 90 K$, or the mechanism is always phonons, and the absence of
the isotope effect in YBCO is due to something else. I prefer the last
option. In its favor speak also the tunneling spectra at higher
voltages in $Nd_{2-x}Ce_xCuO_4$ (NCCO) \cite{[3]} (Fig.2) and in $Bi_2Sr_2CaCu_2O_8$
(BSCCO) \cite{[4]} (Fig.3). The Eliashberg interaction constant
$\alpha^2F(\omega)$
extracted from the tunneling conductance has a shape corresponding to
the phonon density of states $F(\omega)$ obtained from inelastic neutron
scattering. The deep minima in $\alpha^2 F(\omega)$ compared to
$F(\omega)$ can be
attributed to the fact that electrons interact differently with
different phonons. Lately the strong electron-phonon interaction in
BSCCO was confirmed by appearance of branching on the current-voltage
characteristic of a break junction at $eV < 2\Delta $ \cite{[4a]}, representing
phonon-assisted Josephson effect (similar to Shapiro steps) for
low-frequency optical phonons. The frequencies fitted the Raman data
(see sec. 6).

\section{Extended saddle point singularities}

A few years ago two experimental groups: J.-C. Campuzano's from Argonne
\cite{[5]} and Z.-X.Shen's from Stanford \cite{[6]}, investigating
angle resolved photoemission spectra (ARPES), discovered that the
dependence of the energy of quasiparticles on their momentum $\epsilon
(k_x,k_y)$
(the dependence on $k_z$ is weak due to quasi-two-dimensionality) has
flat regions, where it does not depend on one of the components, so
that the spectrum becomes quasi-one-dimensional in these regions
(Fig.4). This was first discovered in YBCO, $YBa_2Cu_4O_8$, BSCCO, and
appeared also in band structure calculations by several groups,
particularly, A.Freeman's \cite{[7]}, for mercury compounds. Such a feature
in the spectrum was called "extended saddle point singularity". The
density of states in the vicinity of such a singularity is
proportional to $(\epsilon - \epsilon_0)^{-1/2}$. If the Fermi energy is
close to $\epsilon_0$, this enhancement is more than sufficient to
explain the high values of Tc assuming the usual strength of the
electron-phonon interaction.  

The second feature is that the integral entering the BCS equation 
\begin{equation}
\Delta({\bf p}) = \int K({\bf p,p'}) f[\Delta({\bf p'})] 
\frac{d^3{\bf p'}}{(2 \pi)^3}
\end{equation}
 which is logarithmically divergent, if the
density of states is constant, becomes convergent. In the case, if
$\epsilon_F - \epsilon_0 \ll \omega_0, \ \omega_0$ being some
characteristic phonon frequency, the critical temperature does not
depend on $\omega_0$, and hence there is no isotope effect. The
BCS-type formula for $Tc$ for this case has the form: 
\begin{equation}
  T_c = \frac{8\gamma(\epsilon_F - \epsilon_0)}{\pi} \exp\Biggl[ -
  \frac{4\sqrt{2}\pi^2(n-1)(\epsilon_F -
    \epsilon_0)}{gm_x^{1/2}\kappa^2}\Biggr]
\end{equation}
The most important here is the appearance of 
$\epsilon_F - \epsilon_0 $ in the exponent and in the prefactor (I
will not explain the meaning of other notations in this expression).
If the distance $\epsilon_F - \epsilon_0$ increases, $T_c$
decreases, and, at the same time, $\epsilon_F - \epsilon_0 $ in the
prefactor is gradually replaced by $\omega_0$ , enhancing the isotope
effect \cite{[8]}. This fits the experimental data.  I would like to stress
that the usual saddle point singularity, extensively studied by
several groups, e.g. J. Friedel, A. Labbe, J. Bok, D. Newns, R.
Markievicz (see \cite{[9]} for references) does not lead to such
consequences. The enhancement of Tc is weak, and no new energy scale
is introduced in the BCS integral, so that the absence of the isotope
effect cannot be explained. Therefore the statement that the flat
regions are equivalent to 2D van Hove singularities is wrong.

\section{Bad Coulomb screening}
 
Under proper assumptions the flat region can dominate in the total
density of states; then, however, we encounter another puzzle. The
ARPES experiments of the same groups established a marked anisotropy
of the energy gap (in the BCS theory the gap corresponds to the
absolute value of the order parameter). The usual belief is that with
phonon-mediated interaction no anisotropy of the order parameter can
be obtained, even in case if the energy spectrum of the metal is
anisotropic. This is due to exchange of phonons with high momenta
which effectively connect all points of the Fermi surface in the BCS
equation (1) (see Fig.5).  However, this statement, being true for
good metals, is not so evident for the substances under consideration.
According to numerous expe-rimental data, they are intermediate
between metals and ionic crystals. The cross-over between these two
limiting cases can be imagined as the change of the Debye screening
radius from interatomic distances to infinity, and the HTSC can be
viewed as substances where all Coulomb-based interactions are screened
at large distances. This refers also to the electron attraction
mediated by phonons, since it is due to the electron-ion interaction.
Therefore it is reasonable to assume this interaction in the form 
\begin{equation}
V({\bf k}) = g \Biggl(\frac{\kappa^2}{k^2 +\kappa^2}\Biggr)^n
\frac{\omega^2(k)}{(\xi-\xi')^2 - \omega^2(k)}
\end{equation}
where $\kappa \ll K$ is the reciprocal Debye radius, $\xi$ is the
electronic energy, and $K$ - the reciprocal lattice period. Since
different models lead to different powers $n$, we will simply assume
$ n > 1$. Under such circumstances the exchange of phonons with small
momenta becomes dominant, and we cannot reach the singular region from
some distant place (Fig.5) without decreasing drastically the value of
the order parameter. Therefore it becomes very anisotropic, and its
maxima are located in the regions with the maximal density of states
\cite{[10]}; this, in fact, is observed in experiment. Moreover the values,
distant from the singular regions, will be defined by the maximal
value, the connection being established through the BCS equation.
Therefore the critical temperature is defined only by these singular
regions.

\section{Two types of the order parameter}

In a purely phonon picture there can be no changes of sign of the
order parameter, and this contradicts many reliable experimental data.
However, apart from phonons, there can exist interactions of some
other origin. The isotope effect tells us that the phonons dominate.
Therefore these other interactions can become important only, if their
nature differs substantially from the phonon-mediated attraction. We
will assume that this additional interaction is a short-range
repulsion coming either from spin fluctuations, or being the
renormalized repulsion at the Cu sites.  To the first approximation
different regions with the maximal density of states and the maximal
absolute value of the order parameter are disconnected. Therefore, the
order parameter can be either even, or odd with respect to a $\pi/2$
rotation in the ab-plane. The corresponding formulas are:
\begin{eqnarray}
\frac{\Delta(\varphi)}{\Delta_1} =  \frac{2(n-1)P_{0y}}{\kappa^2d} \times
\nonumber\\
\Bigg\lbrace \Biggl(\frac{\kappa}{2p_0}\Biggr)^{2n}
\Biggl[\sin^{-2n} \Bigl( \frac{\varphi}{2}\Bigr) + 
\sin^{-2n} \Bigl( \frac{\varphi - \varphi_0}{2}\Bigr) \Biggr]
-\frac{U}{g}\Bigg\rbrace 
\label{even}
\\
\frac{\Delta(\varphi)}{\Delta_1} =  \frac{2(n-1)P_{0y}}{\kappa^2d} 
\Biggl(\frac{\kappa}{2p_0}\Biggr)^{2n} \times \nonumber \\
\Biggl[ \sin^{-2n} \Bigl( \frac{\varphi}{2}\Bigr) -
\sin^{-2n} \Bigl( \frac{|\varphi - \varphi_0|}{2}\Bigr) \Biggr]
\label{odd}
\end{eqnarray}

where $\Delta_1$ is the order parameter in the singular region, 
$P_{0y}$ is the 
length of the singularity, $d$ is the period in the c-direction,
$p_0$  is the radius of the Fermi surface, which we assume cylindrical, 0
and $\varphi_0 = \pi/2$  are the locations of singularities, and $U$ is the
amplitude of the short-range repulsion, which we suppose isotropic.
These expressions are valid far from the singular regions. A simple
interpolation permits to describe the whole Fermi surface. The
simplest formulas for $n=1$ are 
\begin{eqnarray}
\frac{\Delta_{even}}{\Delta_1} =\frac{A}{A + 1 -\cos(4\varphi)},  
\label{even2}
\\
\frac{\Delta_{odd}}{\Delta_1} = \frac{A \cos(2\varphi)}{A + 1 - \cos
(4\varphi)}
\label{odd2}
\end{eqnarray}
(A-adjustable constant), representing the even and odd cases. Fig.6
presents the plot of eq. (\ref{odd2}) for $A=1$. The order parameters 
(4) have the same symmetry, as the"s-wave" and "d-wave" although they
differ from the usual expressions. As already said, to the first
approximation the even and odd states are degeneratre.  This
degeneracy is lifted, if we take into account the energy of the small
overlap of the order parameter originating from different singular
regions. In case, if there were only phonon attraction, the even
solution would have definitely the lowest energy. However, the
presence of a small short-range repulsion with an isotropic Fourier
amplitude in momentum space can drastically change the situation \cite{[11]}.
The interaction Hamiltonian in momentum representation has the shape:
\begin{eqnarray}
H_{int} = - \frac{g}{2} \sum_{\bf p_1 p_2 k} a^\dagger_\alpha ({\bf
  p_1})a^\dagger_\beta ({\bf -p_1 +k}) \times \nonumber \\
V({\bf p_1 - p_2}) a_\beta
  (-{\bf p_2 +k})a_\alpha ({\bf p_2})  \nonumber \\
+ \frac{U}{2} \sum_{\bf p_1 p_2 k} a^\dagger_\alpha ({\bf
  p_1})a^\dagger_\beta ({\bf -p_1 +k}) a_\beta
  (-{\bf p_2 +k})a_\alpha ({\bf p_2}) 
\end{eqnarray}

where $V({\bf k}) =  [\kappa^2/(k^2+\kappa^2)]^n$. Since
the main contribution comes from vicinities of the singular regions,
which are in the a- and b- directions, we separate parts, associated
with the condensate by putting $k=0$ and transforming 
\begin{equation}
a^\dagger_\alpha ({\bf  p_1})a^\dagger_\beta ({\bf -p_1 }) \rightarrow
T \sum_{\omega_n} F^\dagger (\omega_n, {\bf p_1}) I_{\alpha\beta}
\equiv \tilde{F}^\dagger (-{\bf p_1}) I_{\alpha \beta}
\end{equation}

 where $I_{\alpha \beta}$ is the unit antisymmetric matrix:
 $I=i\sigma_y$. 
The $\tilde{F}^\dagger $  can be considered as a creation operator of a Cooper
pair, consisting of electrons with momenta ${\bf p_1}$  and $-{\bf p_1}$.
We leave quadratic and linear terms in $\tilde{F}^\dagger$. 
Then we get 
\begin{eqnarray}
H_{int} \approx -
\frac{2P_{y0}}{(2\pi)^2v_1d}[\lambda_1(\Phi^2_a+\Phi^2_b)
+2\lambda_2\Phi_a\Phi_b] \nonumber \\
- \frac{\lambda_1}{2}[\sum_{\bf p \in a} \Phi_a I_{\alpha\beta} 
(a_\beta({\bf p})a_\alpha (-{\bf p}) + a^\dagger_\alpha({\bf
  -p})a^\dagger_\beta ({\bf p})) +  \nonumber \\ 
\sum_{\bf p \in b} \Phi_b I_{\alpha\beta} 
(a_\beta({\bf p})a_\alpha (-{\bf p}) + a^\dagger_\alpha({\bf
  -p})a^\dagger_\beta ({\bf p}))] \nonumber \\
- \frac{\lambda_2}{2}[\sum_{\bf p \in b} \Phi_a I_{\alpha\beta} 
(a_\beta({\bf p})a_\alpha (-{\bf p}) + a^\dagger_\alpha({\bf
  -p})a^\dagger_\beta ({\bf p}))   \nonumber \\ 
- \sum_{\bf p \in a} \Phi_b I_{\alpha\beta} 
(a_\beta({\bf p})a_\alpha (-{\bf p}) + a^\dagger_\alpha({\bf
  -p})a^\dagger_\beta ({\bf p}))]
\label{h}
\end{eqnarray}
where
\begin{eqnarray}
\lambda_1 = \frac{1}{(2\pi)^2v_1} \Biggl( \frac{g\kappa^2}{n-1} -
\frac{2UP_{y0}}{d}\Biggr), \nonumber \\
 \lambda_1 = \frac{2P_{y0}}{(2\pi)^2v_1}
\Biggl[ g\Biggl( \frac{\kappa^2}{2p_0^2} 
\Biggr)^n - U \Biggr] \\
\Phi_{a,b} =
\int^\infty_{-\mu_1}\Biggl( \frac{\mu_1}{\xi+\mu_1}\Biggr)^{1/2}
\tilde{F}_{a,b}(\xi)d\xi 
\label{(9)}
\end{eqnarray}
We used here that in the singular region the functions 
$\tilde{F}_{a,b}$  depend only on the normal component of the
momentum, i.e., on $\xi =  v_1 (p_n - p_0)$, and performed integrations over
other components. Since we assume that $g\kappa^2 \gg UP_{y0}/d$
and $\kappa \ll p_0$, we have: $\lambda_1 \gg \lambda_2 $, and hence the
overlap terms in the Hamiltonian are indeed small. We would like to
stress here that the reduced interaction Hamiltonian (\ref{h}), where only
the singular momentum regions were left, is sufficient for calculation
of the critical temperature and total energy but fails when the
low-energy excitations play the major role, as in the specific heat
and kinetics at low temperatures.  Applying all the superconducting
machinery (see ref.\cite{[12]}) to the full Hamiltonian, including kinetic
terms, we obtain the necessary quantities: 
\begin{eqnarray}
T_{c, \,{\rm odd}} = 8 \frac{\gamma}{\pi}\mu_1
\exp[-1/(\lambda_1-\lambda_2)] \nonumber \\ 
= \frac{\gamma}{\pi}\Delta_1
\exp\{-\lambda_2/[\lambda_1(\lambda_1-\lambda_2)] \} \nonumber \\
\approx \frac{\gamma}{\pi}\Delta_1 \exp(-\lambda_2/\lambda_1^2) \nonumber \\
T_{c, \,{\rm even}} \simeq \frac{\gamma}{\pi}\Delta_1
\exp(\lambda_2/\lambda_1^2), \lambda_2 \ll \lambda_1  \nonumber \\
T_{c, \,{\rm odd}} - T_{c, \,{\rm even}} \approx - 2
\frac{\gamma}{\pi}\Delta_1 \sinh(\lambda_2/\lambda_1^2), \nonumber \\
 \Delta_1 =
8 \mu_1 \exp(-1/\lambda_1) 
\label{(10)}\\
\Omega_{\rm odd} (0) - \Omega_{\rm even} (0) \approx 
\frac{4P_{y0}}{(2\pi)^2 v_1 d}\Delta^2_1 \sinh(\lambda_2/\lambda_1^2)
\end{eqnarray}
From (\ref{(9)}) and (\ref{(10)}) it can be seen that, if $\lambda_2 >
0$, the equilibrium solution is even, and, if $\lambda_2 < 0$ , it is
odd. This occurs at all temperatures, where superconductivity exists.
From the definitions (\ref{(10)}) it follows that the phonons dominate in the
interaction, if 
\begin{equation}
g\kappa^2/(n-1) \gg 2UP_{y0}/d
\end{equation}
Under this condition $\lambda_2$ can have either sign. It is negative,
if 
\begin{equation}
U > g [\kappa^2/(2p_0^2)]^n
\end{equation}
The main conclusion is the division of layered cuprates in two very
distinct groups. Strong anisotropy, originating from extended saddle
point singularities and a long screening radius, makes it plausible
that the order parameter becomes odd, whereas weak anisotropy leads to
an even order parameter.  Since the quasi-1D energy spectrum leads to
a singularity in the density of states only from one side, it is
likely to be absent in the n-type cuprates, as NCCO . This can be the
origin of their lower $T_c$  and of the even order parameter.

\section{Scattering from impurities}

A strong proof of the odd order parameter is the action of impurities
\cite{[13]}. Regular substitutional impurities are most likely ionized, and
being small-angle scatterers, they do not influence $T_c$
substantially. Of course, there is also some large-angle scattering
but it has a small probability. The usual estimates lead to the
conclusion that the critical temperature is influenced only if 
$\tau\epsilon_F \sim 1$, which could be considered in favor of
s-pairing. The fault of such an estimate is that a much larger $\tau$ 
should be taken, namely, that of large angle scattering, and it could
fit the condition $\tau\Delta \sim 1$. In 
the experiment \cite{[14]}, defects were created in YBCO by electron
bombardment. The electron energy was chosen so that the oxygen atoms
were displaced but not knocked out; this produced dipolar defects
causing large-angle scattering. They were very effective in
suppression of the critical temperature. At the same time the copper
atoms were not involved, and this excluded magnetic defects.  The
observed dependence of Tc on the concentration of these defects fits
the prediction for an odd order parameter: 
\begin{eqnarray}
T_{\rm co} \approx \left\{ \begin{array}{ll}
T_{co}^{\rm clean} [1 - \pi/(2T_{co}^{\rm clean} \tau)], & (\pi
T_{co}^{\rm clean} \tau )^{-1} \ll 1 \\
\frac{2\sqrt{6}}{\pi} \sqrt{\tau_c^{-1} (\tau_c^{-1} - \tau^{-1})}, &
(\tau^{-1}_c - \tau^{-1}) \ll \tau_c^{-1}
  \end{array}
\right. 
\end{eqnarray}
where $\tau^{-1}$ is the scattering probability proportional to the
residual resistance in the normal state, and 
\begin{equation}
\tau_c^{-1} = (\pi/4\gamma) T_{co}^{\rm (clean)}
\end{equation}
is the critical value, where $T_{co}$ 
vanishes.  In reality, however, this does not happen, because the
metal can undergo a phase transition into the even phase. This was not
observed in the experiment \cite{[14]} in the accessible temperature
interval, however there exist other ways to suppress the d-wave order
parameter, and at least in one of such experiments performed by the
group of L.  Greene a "subdominant" s-wave order parameter was indeed
observed \cite{[15]}.

\section{Negative isotope effect on copper}

Rather close considerations can explain the negative isotope effect on
copper, discovered by the Franck's group \cite{[16]}. It was found that in
underdoped samples of YBCO the substitution $^{63}Cu - ^{65}Cu$ causes a
change of $T_c$ corresponding to a negative isotope effect, namely, a
larger copper isotope mass leads to an increase of the critical
temperature. There exist some other examples of a negative isotope
effect, e. g., resulting from a substitution of hydrogen by deuterium
in solid solutions. The distinction of the present case is that the
same substances have a positive isotope effect for the oxygen
substitution $^{16}O - ^{18}O$. The absolute value of the negative copper
isotope effect depends on the oxygen concentration. In optimally doped
samples it is small, otherwise it depends nonmonotonically on
concentration. The maximal absolute value is around 0.4; we have in
mind the power a in the dependence $Tc \propto M^{-\alpha}$, or for small
variations of M, 
\begin{equation}
\alpha = -\Delta(\ln T_c)/\Delta(\ln M).
\end{equation}
 In a simple model one
cannot hope to reproduce all the details, and our goal will be only
the qualitative explanation of the most spectacular features. All
phonons, whatever their nature is, contribute to the attractive
interaction, and therefore, in order to explain the negative isotope
effect, we have to think about some different action. In this
connection we may remember that in a d-wave superconductor ordinary
potential nonmagnetic scatterers with large angle scattering suppress
superconductivity. Let us suppose that there exist phonons which do
not affect strongly the overall ionic charge. The interaction of
electrons with these phonons will be short-ranged, the same as with
neutral defects. If we suppose also that the characteristic frequency
of such phonons is small compared to $Tc$, we obtain a situation similar
to elastic impurity scattering. In a d-wave superconductor this
scattering will suppress superconductivity, and its effect will
increase with increasing amplitude of oscillations, i. e. with
decreasing mass of the ions involved. This consideration is indeed
justified by exact calculations. The result is 
\begin{equation}
\alpha \approx - 6.326 g_2^2 \nu \Bigl( \frac{\omega_{02}}{2\pi T_c}\Bigr)^2,
\end{equation}

where $\omega_{02}$ and $ g_2^2$ are the frequency of Cu phonons with
momentum $p_0\sqrt{2}$ and their interaction constant with electrons,
$\nu $ is the density of states. The effect increases in magnitude with
decreasing $T_c$, i. e. with underdoping. For an s-type order
parameter the isotope effect is always positive.  The Raman scattering
experiments \cite{[17]}, \cite{[18]} show that indeed low frequency
phonons do exist. The best candidate is the mode with $\omega_0
\approx 150 cm^{-1}$ which corresponds to vibrations of Cu atoms. For
this mode the expansion parameter of the theory $(\omega_0/2\pi T_c)^2
 \sim 0.24 $
(we took $T_c = 70 K$).

\section{Neutron peak}

Another interesting consequence of the proposed theory is the
explanation of the origin and the prediction of the fine structure of
the peak in inelastic neutron scattering in YBCO \cite{[19]}. The clear
narrow peak was observed in the superconducting phase for momentum
change close to $(\pi/2,\pi/2)$ and energy change close to 41 meV
(\cite{[20]}, and references thereon), which corresponds, approximately, to
$2\Delta_{max}$, as obtained from tunneling measurements \cite{[21]}. The
momentum change corresponds to the transition from one extended saddle
point singularity to another one. In these regions the spectrum is
almost one dimensional.  At Fig.7 the energy in the normal state is
presented as function of the corresponding momentum: $p_x$  for the
a-vicinity, and $p_y$ for the b-vicinity (dashed curve). In order
to obtain from here the superconducting spectrum one has to reflect
the curve with respect to the chemical potential (straight horizontal
line) and introduce gaps at the intersections. Then we get the two
continuos curves at Fig.7. The momentum and energy lost by the neutron
are absorbed by the electron system and can be described, as
transitions from the lower to the upper curve. Singular points in the
scattering probability as function of energy can appear, if both, the
initial and the final point correspond to the maxima of the density of
states, i.e. to the extrema of the curves. The lowest energy
corresponds to the transition between the maximum of the lower curve
(1) to the minimum of the upper (1'), and at $T=0$ this is the
threshold, located at 2D. Then comes the transition from the maximum
of the lower curve (1) to the maximum of the upper (2'). Ths must be a
maximum but it is not a threshold, since processes with smaller and
larger energies are possible. Finally comes the transition from the
minimum of the lower curve (2) to the maximum of the upper (A'). This
corresponds to the end of an absorption path, and hence the scattering
probability must have a drop at this point. Rigorous calculations
confirm this picture. The result is presented at Fig.8. This has to be
compared with the experimental curve at Fig.9 \cite{[19]}.  It should be
added that due to the so called "coherence factor" the maximum appears
only, if the order parameter has different signs for the initial and
final states (d-wave symmetry). This makes the inelastic neutron
scattering, as well as the negative isotope effect described above
rather unique bulk measurements confirming the d-wave order parameter.
The measurements associated with surfaces and interfaces can reveal
properties different from the bulk, as, e.g., the suppression of the
d-wave order parameter and appearance of a subdominant s-wave.
Another important feature is that in a two-layered substance, which is
YBCO, the maximum appears only in the channel which corresponds to the
interaction of neutrons with the electron spin fluctuations
antisymmetric with respect to the layers. Such fluctuations correspond
to "acoustical" gapless spin-wave branch in the antiferromagnetic
insulating phase. This is natural, since other fluctuations
immediately decay into quasiparticles. On the other hand, such
fluctuations are rather long ranged, and therefore they define,
actually, the true "coherent" gap. It is no surprise, therefore, that
with underdoping the position of the neutron maximum varies
proportional to Tc and not to the gap in the quasiparticle spectrum
measured in tunneling experiments.

\section{Upper critical field}

The last interesting consequence of the model, which I will describe,
is the positive curvature of the upper critical magnetic field, as
function of temperature \cite{[22]} which was observed in many experiments,
(e.g. \cite{[23]}, see Fig.10). The usual calculation gives a negative
curvature. The reason for such a change is contained in the form of
the spectrum. As it was mentioned before, the most important are the
vicinities of the singular regions, where the motion of electrons
becomes one-dimensional. There is, however, a hopping between the "a"
and "b" vicinities, and hence a two-dimensional motion. Therefore,
close to $T_c$ the field $H_{c2}$  is defined by the usual orbital
motion. With departing from  Tc the situation changes. This can be
seen from the Ginsburg-Landau type equations derived in the vicinity
of $T_c$. They correspond to the minimum of the free energy
\begin{eqnarray}
F_0 =  \int dx\,dy\Bigl\{ \frac{1}{4m_x} \Bigl[ |\Bigl(-i
\frac{\partial \,}{\partial x} -  \frac{2e}{c} A_x \Bigr) \Psi_a|^2 
\nonumber \\
+ |\Bigl(-i
\frac{\partial \,}{\partial y} -  \frac{2e}{c} A_y \Bigr) \Psi_b|^2\Bigr]
\nonumber \\
+ \alpha [\tau(|\Psi_a|^2 + |\Psi_b|^2) + \eta |\Psi_a + \Psi_b|^2 
\nonumber \\
+ (1/2n) (|\Psi_a|^4 + |\Psi_b|^4) ] + 
\frac{H^2d}{8\pi}\Bigr\}
\end{eqnarray}
where
\begin{equation}
\alpha = \frac{2(\pi T_c)^2}{7\zeta(3) \mu_1}, \, \tau = \frac{T -
  T_c}{T_c}, \, \eta = \frac{|\lambda_2|}{(\lambda_1 + |\lambda_2|)^2}
\end{equation}

The chemical potential $\mu_1$ is calculated with respect to the
bottom of the band. The term with $\alpha\tau$ is negative below $T_c$
and it keeps $\Psi_{a,b}$ attached to a certain vicinity, whereas the
term with $\alpha\eta$ corresponds to hopping between them. With
increasing $|\tau|$ the motion becomes more one-dimensional, and eventually
the orbital motion is unable to destroy superconductivity, i.e.,
$H_{c2}$. For sufficiently small values of h this can happen even
close to Tc (see Fig.11). In the immediate vicinity of Tc
\begin{equation}
H_{c2} \approx \frac{4mc}{e} \alpha |\tau| \Bigl( 1 + \frac{|\tau|}{\eta}\Bigr)
\end{equation}
For sufficiently high fields another mechanism becomes important,
namely the paramagnetic destruction of Cooper pairs.  Unfortunately,
the account of this effect leads to the restoration of a negative
curvature at lower temperatures and larger fields. This does not fit
the experimental data (see \cite{[24]}).
      
There was an interesting suggestion \cite{[25]} that in the cases of reduced
dimensionality (films, wires in longitudinal field), where the orbital
destruction becomes impossible, the paramagnetic effect leads to a
sequence of phase transitions between different inhomogeneous LOFF
(Larkin-Ovchinnikov-Fulde-Ferrell) phases with increasing critical
fields. This is a possible origin of the positive curvature at very
low temperatures.

\acknowledgments
        
This work was supported by the Department of Energy under the
contracts \# W-31-109-ENG-38.

\begin{figure}
\centerline{\includegraphics[width=1.75in]{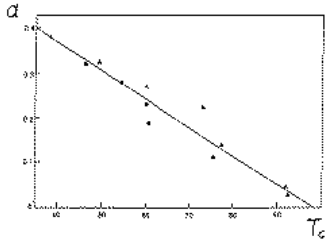}}
\caption{Experimental results for the isotope shift ($O_{16}
  \rightarrow O_{18}$) in 
  partially substituted YBCO ($Y\rightarrow Pr$, and $Ba \rightarrow
  La$). The exponent $\alpha$ in $T_c \propto M^\alpha$ is presented
  as function of $T_c$ together with 
  the linear regression.}
\end{figure}

\begin{figure}
\centerline{\includegraphics[width=2.25in]{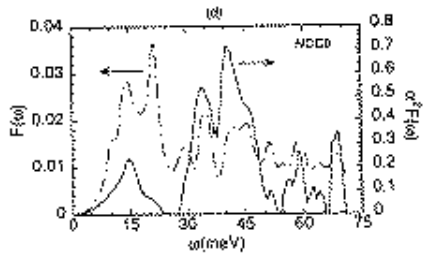}}
\caption{Compilation of tunneling $\alpha^2F(\omega)$ for NCCO (solid line)
  and the phonon density of states $F(\omega) $
  generated from the dispersion curves obtained from the inelastic
  neutron scattering from single-crystal NCCO}
\end{figure}

\begin{figure}
\centerline{\includegraphics[width=1.5in]{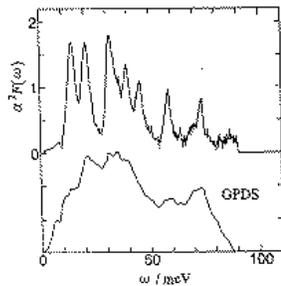}}
\caption{The same, as in
  Fig. 2, for BSCCO. The lower curve is $F(\omega)$ from inelastic neutron
  scattering}
\end{figure}

\begin{figure}
\centerline{\includegraphics[width=1.75in]{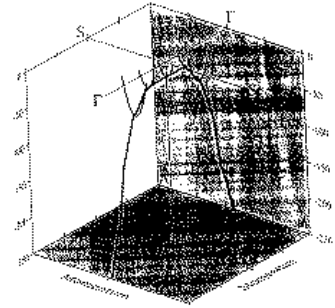}}
\caption{ARPES result for the dependence of quasiparticle
  energy on components of quasimomentum in the vicinity of the
  extended saddle point singularity.}
\end{figure}

\begin{figure}
\centerline{\includegraphics[width=1.75in]{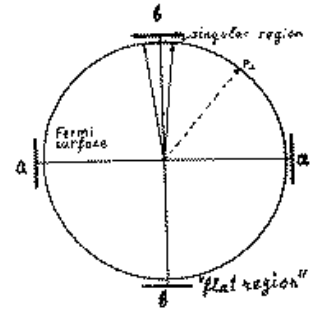}}
\caption{Fermi surface with
  singular regions having a high density of states. A short range
  interaction connects any point with the singular region, whereas a
  long range interaction connects only points within this region.}
\end{figure}

\begin{figure}
\centerline{\includegraphics[width=2in]{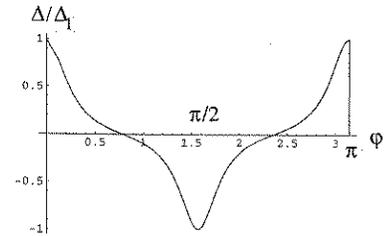}}
\caption{Plot of $\Delta/\Delta_1$ according to
  eq. \protect\ref{odd}.} 
\end{figure}

\begin{figure}
\centerline{\includegraphics[width=2in]{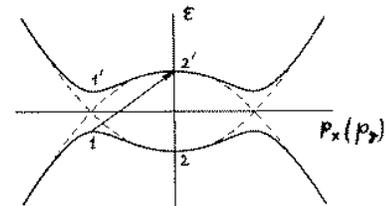}}
\caption{Schematic presentation of the superconducting bands in the vicinity
  of singular points. The abscissa is $p-x$ for the a-vicinity,
  and $p_y$ for the b-vicinity. The notations 1,1',2,2' correspond
  to extremal points with the maximal density of states.}
\end{figure}

\begin{figure}
\centerline{\includegraphics[width=2in]{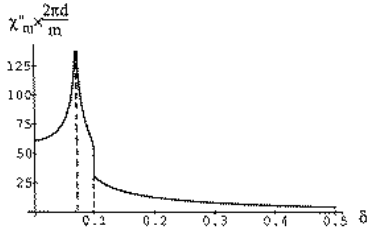}}
\caption{Plot
  of the imaginary part of the electronic spin susceptibility, which
  is the energy dependent factor in the neutron inelastic scattering
  cross section $\chi_{m}''\times \frac{2\pi d}{m}$ (m-effective mass,
  d-period along), as function of $\delta \equiv (\Omega/\Delta
  -2)^{1/2}$ where $\Omega$ is the energy loss, in the case $\gamma
  \equiv \mu_1/\Delta = 0.1$ ($\mu_1$ is the Fermi energy with respect
  to the bottom of the 1D band). Here the point $\Omega = 2 \Delta$ is
  the threshold, and below this poind $\chi'' =0$ }
\end{figure}

\begin{figure}
\centerline{\includegraphics[width=2in]{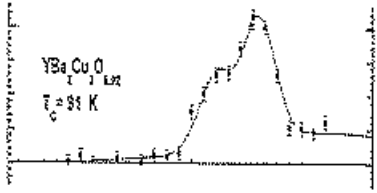}}
\caption{Experimental data from \protect\cite{[20]} for the same quantity in
  arbitrary units as function of energy loss for optimally doped YBCO.
  The initial rise corresponds to the smeared out threshold.}
\end{figure}

\begin{figure}
\centerline{\includegraphics[width=2in]{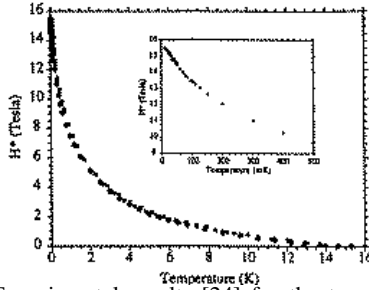}}
\caption{Experimental results \protect\cite{[23]} for the temperature
  dependence of the upper critical field}
\end{figure}

\begin{figure}
\centerline{\includegraphics[width=2in]{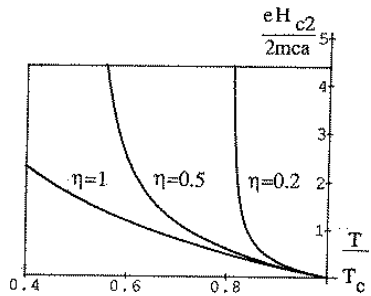}}
\caption{Theoretical dependence
  of $eH_{c2}/(2mc\alpha)$ on $T/T_c$ for different values of $\eta$.}
\end{figure}
\end{document}